\documentstyle[12pt]{article}
\begin{document}

\title{
Unitarity Constraints for the Mass of the Higgs in the
$SU(2)_L\otimes U(1)\otimes U(1)'$ Gauge Model}
\author{R. Mart\'{\i}nez\thanks{Also Universidad Nacional, Depto.
de F\'{\i}sica, A.A. 14490, Bogot\'a, Colombia} \\
CINVESTAV \\
Depto. de F\'{\i}sica \\
Apartado Postal 14740 \\
07000 M\'exico \\
\and\\
N. Vanegas\thanks{Supported by Colciencias and U. de Antioquia,
A.A. 1226, Medell\'{\i}n, Colombia.}  \\
Physics Department,\\
Queen Mary and Westfield College, \\
Mile End Road, London E1 4NS\\
{\em email:} n.vanegas@qmw.ac.uk}
\date{}
\maketitle
\noindent {\bf Abstract} \\
The critical values for the mass of the Higgs bosons, at which
the theory becomes strongly interacting, are calculated using the
equivalence theorem, at high energies this allows us to replace
the longitudinally polarized gauge bosons in the $S$ matrix for
the corresponding Goldstone bosons. An appropriate ansatz for
defining the would-be Goldstone bosons in the case of an
additional neutral current, beyond the minimal standard model, is
also presented.

\newpage

The minimal standard model (MSM) gives an adequate description of
the available experimental data on the electroweak interaction,
but, theorists believe, there is physics beyond the MSM and hope
that the advent of new accelerators will show some experimental
evidence of this. This new physics should bring the need for new
models like the supersymmetric models, technicolour or simple
extensions to the gauge group. One can think in many different
extensions to the gauge group, some of the most relevant are:
the modification of the Higgs sector (including
two  $SU(2)_L$ Higgs doublets, one additional Higgs singlet, one
additional $SU(2)_L$ triplet or more complicated Higgs
structures \cite{triplet}), and the use of
richer gauge groups, left-right symmetric models
$SU(2)_L\otimes SU(2)_R \otimes U(1)_{B-L}$ \cite{derecho}, or
additional $U(1)$ groups $SU(2)_L \otimes U(1)
\otimes U(1)'$ \cite{adicion}. However, the origin of the
spontaneous symmetry breaking, needed to generate the gauge boson
masses, remains unclear.

In the MSM with one Higgs doublet \cite{salam} there are three
would-be Gold~stone bosons `eaten' by the gauge fields,
which become massive  in the process  and leave one neutral
scalar
particle: the Higgs boson. The  current experimental lower bound
for the mass ($m_H$) of this boson stands at about 60
GeV~\cite{lore}. The actual value of $m _H$ is crucial  for the
validity
of the MSM at high energies, recalling
the fact that the unitarity  of the theory is not preserved at high
energies
if $m_H$ exceeds a critical value of about 1 TeV  \cite{georgi}. In
this case the scalar sector of the MSM
becomes strongly interacting and the perturbative
expansion of the $S$ matrix is no longer valid, thus chiral
\cite{chiral} or effective \cite{efectivo} Lagrangian approaches
may be an appropriate description for the gauge bosons physics
since they only use the symmetry breaking scheme.

This can be seen if we consider  the high energy  limit (compared
with the mass scale of the particles involved). The polarization
vector of a vector boson is
\begin{equation}
\epsilon^{\mu} = (\frac{{\bf k} \cdot {\bf \epsilon}}{M} , {\bf
\epsilon} + \frac{{\bf k}({\bf k} \cdot {\bf \hat{\epsilon}})}{M
(Q + M)})
\end{equation}
with $K = (Q, {\bf k})$ and $K^2 = M^2$, where it is easy to see
that, at high energies, the dominant part of the gauge bosons is
that of longitudinal polarization which, in the same limit, can be
written as
\begin{equation}
\epsilon^{\mu} \simeq \frac{K^{\mu}}{M} \, .
\end{equation}
Moreover, in the 't-Hooft - Feynman gauge (a special case of the
so called
$R_{\xi}$ gauges which we will discuss later in a more general
form), the gauge fixing term is given by
\begin{equation}
\partial_{\mu} V^{\mu}_{i} + i M \phi_{i} = 0 ,
\end{equation}
where $V_{i}$ and $\phi_{i}$ are any vector boson field and its
corresponding Goldstone boson, one can replace $V_i$ with
$\phi_i$ in any $S$ matrix calculation at high energies,
\begin{equation}
S[ V_i ] = S [ \phi_i ] + S[ \cal O (M/Q) ]
\end{equation}
therefore the
scattering amplitudes for longitudinal $W$'s and $Z$'s can be
calculated from the scattering amplitudes of the would-be
Goldstone bosons \cite{boson}, up to order $M/Q$ (only in this
gauge);
this is known as the Equivalence Theorem, proposed by Lee, Quigg
and Thacker is 1977.

The main result of this paper is to present an ansatz for
obtaining the corresponding
would-be Goldstone bosons in the $SU(2)_L\otimes U(1)_Y \otimes
U(1)'$ model, with one $SU(2)$ doublet and two singlets. We also
find the upper bound for the mass of the Higgs bosons in the
limit $\sqrt{s} \gg m_H$  by looking at the scattering process of
two neutral
Goldstone bosons. Via the equivalence theorem this is a good
approximation for the scattering amplitude of two neutral high
energy
gauge bosons, with longitudinal polarization. Finally, we find a
relation
between the mass of the extra Higgs boson, the new neutral gauge
field and  the mixing angle between the neutral currents.
\vspace{3mm}

The $SU(2) \otimes U(1) \otimes U(1)'$  model is important
because future experiments may find
additional neutral currents and, even if it is not the exact
gauge group of the electroweak interaction, it provides the
minimal model with such an extra current, therefore it might be
useful when considering some superstring theory-based models,
specially
those with $E_6$ symmetry or models with left-right symmetry
mentioned above. At low
energies these models might behave like $SU(2)_L \otimes U(1)
\otimes U(1)'$ \cite{aguila}.

The most general expression for the electric charge in
$SU(2)_L\otimes U(1)\otimes U(1)'$ is
\begin{equation}
Q = T_{3L} + (aY_1+bY_2)/2,
\end{equation}
where $T_{3L}$, $Y_1$ and $Y_2$ are the diagonal generators of
$SU(2)_L$, $U(1)$ and $U(1)'$ respectively. The second term is
$aY_1+bY_2=Y_{GWS}$ for a given multiplet, where $Y_{GWS}$ is the
hypercharge in the Weinberg-Salam model. The most general
Lagrangian for the bosonic sector in this model with one $SU(2)$
doublet and one singlet is given by
\begin{eqnarray}
L &=&\sum_{i=1,2} \left(
\; (D_{\mu} \Phi_{i})^{\dagger} (D^{\mu} \Phi_{i}) +
\lambda_i \; [ \; \Phi^{\dagger}_{i} \; \Phi_{i} - v^{2}_{i} \;
]^{2} \; \right) \nonumber \\
&+&\lambda_3 (\Phi^{\dagger}_1 \Phi_1) (\Phi^{\dagger}_2 \Phi_2),
\end{eqnarray}
where the Higgs field components are:
\begin{eqnarray}
\Phi_1 &=& \left(
\begin{array}{c}
(\varphi_1 + i \varphi_2)/ \sqrt{2}  \\
(H + i \varphi_3)/ \sqrt{2} \nonumber
\end{array}
\right ) \; \; , \nonumber  \\
\Phi_2 &=& (\chi + i \varphi_4)/ \sqrt{2} \; ,
\end{eqnarray}
and the covariant derivative is \cite{ponce}
\begin{equation}
D_{\mu} = \partial_{\mu} - ig \vec{T}. \vec{A_{\mu}} -
i\frac{g_1}{2} Y_1 B_{\mu} - i\frac{g_2}{2} Y_2 C_{\mu}  .
\end{equation}

The Higgs fields have vacuum expectation values
\begin{eqnarray}
<H>_o &=& v_1  \; , \nonumber \\
<\chi>_o &=& v_2,
\end{eqnarray}
and then the symmetry is spontaneously broken to $U(1)_Q$ .

The mass matrix obtained from the Higgs potential for the fields
$\varphi_3$, $\varphi_4$ (the neutral would-be
Golstone bosons) is identically zero, and it is unknown which
of the fields are `eaten' by the longitudinally polarized
gauge fields. To get the renormalizable theory, in the $R_{\xi}$,
gauge we need to cancel the mixing terms
$\eta_{Z}\left(\partial_{\mu} Z^{\mu} \right)$, $\eta_{Z^{'}}
(\partial_{\mu}Z^{\mu '})$ and $\varphi^+ \left(\partial_{\mu}
W^{\mu -} \right)$ from the kinetic lagrangian of the Higgs
fields, where $\eta_Z$, $\eta_{Z'}$ and $\varphi^+$ are the
Goldstone bosons of the $Z_{\mu}$, $Z_{\mu}'$ and
$W_{\mu}^+$, respectively. The gauge fixing term in the
lagrangian is \cite{gauge}
\begin{eqnarray}
L_{GF} = &-&\frac{1}{2 \xi_{a}} \left( \partial_{\mu} A^{\mu}
\right)^{2} -
\frac{1}{2 \xi_{z}} \left( \partial_{\mu} Z^{\mu}
 - \xi_{z} M_{Z} \eta_{Z} \right)^2 \\
 &-& \frac{1}{2 \xi_{z'}} \left(\partial_{\mu} Z^{\mu '} -
\xi_{z'}  M_{Z'} \eta_{Z'} \right)^2
- \frac{1}{ \xi_{w}} \left| \partial_{\mu}
W^{\mu +} - i \xi_{w} \; M_{W} \; \varphi^{+} \right| ^2 \, ,
\nonumber
\end{eqnarray}
and, using the covariant derivative as a function of the real
fields \cite{ponce}, we get
\begin{eqnarray}
\eta_Z &=&  \frac{g}{M_Z} \frac{\sin\psi}{\cos\theta}
\; \sum_{i=3,4} v_i  \left(t_{3i}
-\frac{\sin\theta \cot\psi}{ \sin 2\xi } [ a
Y_{1\varphi_{i}} - 2Y_{GWS_{i}} \sin^2\xi] \right) \varphi_i
 \; , \nonumber \\
\eta_{Z'} &=& \frac{g}{M_{Z'}} \frac{\cos\psi}{\cos\theta}
\sum_{i=3,4}v_i
\left(t_{3i}+\frac{\sin\theta\tan\psi}{ \sin 2\xi }
[aY_{1\varphi_{i}} - 2Y_{GWS_{i}}\sin^2\xi] \right) \varphi_i
\end{eqnarray}
where $M_Z$ and $M_{Z'}$ are the $Z_{\mu}$ and $Z_{\mu}^{'}$
masses respectively and $t_{3i}$ are the
$T_{3L}$ quantum numbers of the Higgs fields. The angles
$\theta, \psi$ and $\xi$ are given by
\begin{eqnarray}
\tan \xi &=& \frac{a}{b} \; \frac{g_2}{g_1} \; , \nonumber \\
\sin\theta &=& \frac{e}{g} \; , \nonumber \\
\tan 2\psi &=& - \frac{4 M_W^2 / \cos^2
\theta}{M_{Z'}^2+M_{Z}^2-2M_W^2/ \cos^2 \theta} \; \frac{\sin
\theta}{\sin^2 \xi} \; ( aY_{1} - \sin^2 \xi )
\end{eqnarray}
where $g1$, $g2$ and $e$ are the coupling constants associated
with $U(1)$, $U(1)'$ and $U(1)_Q$ respectively, $\theta$ is the
Weinberg angle and the angle $\psi$ gives the mixing between the
weak neutral currents. If $\psi
= 0 $ there is no mixing and we get the SM with one additional
multiplet plus one extra term due $Z'$.
According to some recent results from LEP and the L3 collaboration data
\cite{lep} the
possible values of the mixing angle are in the range
$\left|\psi\right| \leq 0.03 - 0.01$.

{}From Eq.\ (7) and the quantum numbers of the Higgs multiplets,
\begin{eqnarray}
\varphi_3 &=& \alpha \, \eta_Z + \beta \, \eta_{Z'}, \nonumber\\
\varphi_4 &=& \rho \, \eta_Z + \sigma \eta_{Z'}
\end{eqnarray}
in the limit when the mixing angle between $Z_{\mu}$ and
$Z'_{\mu}$ is  $\psi \approx 0$, we have:
\begin{eqnarray}
\alpha&=&-\frac{2\cos\theta \cos\psi M_Z}{gv_1}
\approx - 1 \; , \nonumber  \\
\rho&=&\frac{2\cot\theta M_Z}{g a
Y_{1\phi_{4}}v_2}\left[-\sin\xi\cos\xi \sin\psi +
\sin\theta\cos\psi(aY_{1\phi_{3}}-2\sin^2\xi)\right] \nonumber\\
&\approx& \frac{M_Z \sin\psi}{M_{Z'}} \; , \nonumber \\
\beta&=&-\frac{2\cos\theta \sin\psi M_{Z'}}{g v_1}
- \approx \frac{M_{Z'}\sin\psi}{M_Z} \; , \nonumber \\
\sigma&=&\frac{2\cot\theta M_{Z'}}{g a
Y_{1\phi_{4}}v_2}\left[\sin\xi\cos\xi\cos\psi+\sin\theta
\sin\psi(aY_{1\phi_{3}}-2\sin^2\xi)\right] \nonumber\\
&\approx& \cos\psi
\end{eqnarray}

Now let us consider the processes $\eta_{Z}  \eta_{Z} \rightarrow
\eta_{Z} \eta_{Z}$ and $\eta_{Z'} \eta_{Z'} \rightarrow \eta_{Z'}
\eta_{Z'}$, the scattering amplitudes are:
\begin{eqnarray}
T(\eta_Z \eta_Z \rightarrow \eta_Z\eta_Z)=&-&\frac{i\alpha^4
m_H^2}{v_1^2}  \; \left[\frac{s}{s-m_H^2}+\frac{t}{t-m_H^2} +
\frac{u}{u-m_H^2} \; \, \right] \nonumber\\
&-& \, \frac{i\rho^4 m^2_{\chi}}{v_2^2} \, \;
\left[\frac{s}{s-m_{\chi}^2} \, +
\frac{t}{t-m_{\chi}^2} \, +\frac{u}{u-m_{\chi}^2} \, \; \;
\right] \; , \nonumber \\
T(\eta_{Z'} \eta_{Z'} \rightarrow
\eta_{Z'}\eta_{Z'})=&-&\frac{i\sigma^4m_{\chi}^2}{v_2^2} \; \;
\left[\frac{s} {s-m_{\chi}^2} \, +\frac{t}{t-m_{\chi}^2} \,
+\frac{u}{u-m_{\chi}^2} \, \; \; \right] \nonumber \\
&-& \, \frac{i\beta^4m_H^2}{v_1^2} \, \left[\frac{s}{s-m_H^2} \,
+\frac{t}{t-m_H^2} \, +\frac{u}{u-m_H^2} \, \right],
\end{eqnarray}
where $m_H, m_{\chi}$ are the neutral Higgs bosons masses.
Recalling  that the mixing angle between the neutral gauge bosons
is small and for
simplicity we assume that the mixing between $H$ and $\xi$
is zero, i.e. $\lambda_3=0$.

The scattering  amplitude can be decomposed in partial waves,
according to
\begin{equation}
T(s, \theta) = 16 \pi  \sum_{j=0}^{\infty}  \, a_j \, P_j(\cos
\theta) \, ,
\end{equation}
at high energy ($s>\!>m_H^2, m_{\chi}^2$) the tree-level
contributions  to the $j=0$ partial wave amplitudes are given by
\begin{eqnarray}
a_0(\eta_Z\eta_Z\rightarrow\eta_Z\eta_Z) & = &
-\frac{3i}{8\sqrt{2}\pi} \left[\frac{\alpha^4m_H^2}{v^2_1} +
\frac{\rho^4m_{\chi}^2}{v^2_2}\right] \, , \nonumber \\
a_0(\eta_{Z'}\eta_{Z'}\rightarrow\eta_{Z'}\eta_{Z'}) & = &
-\frac{3i}{8\sqrt{2}\pi} \left[\frac{\sigma^4m_{\chi}^2}{v^2_2} +
\frac{\beta^4m_H^2}{v^2_1}\right] \; .
\end{eqnarray}
{}From unitarity requirement the upper bound for the mass of
the Higgs is therefore,
\begin{eqnarray}
m_H^2 &\leq& \frac{4\pi v_1^2}{3\alpha^4}\approx
\frac{4\pi\sqrt{2}}
{3G_F\cos^4\psi} \; , \nonumber \\
m_{\chi}^2 &\leq& \frac{4\pi v_2^2}{3 \sigma^4}\approx \frac{4\pi
v_2^2} {3\cos^4\psi} \nonumber \\ &\leq& \frac{4\pi\sin^{2}2\xi
M_{Z'}^2}{3\sqrt{2}G_{F}\sin^2\theta M_{Z}^2 \cos^4\psi}\; ,
\end{eqnarray}
where we have made use of the mass relation for $M_{Z'}$ obtained
in Ref.\cite{ponce}, given by
\begin{equation}
M_{Z'}^2=\frac{g^2 v_2^2 \tan^2\theta}{4 \sin^2 2\xi} \; .
\end{equation}

An interesting case occurs when one considers the limit $m_{H}
\leq m_{\chi}$,
\begin{equation}
m_{H}^2 \leq m_{\chi}^2 \leq \frac{4\pi\sin^{2}2\xi M_{Z'}^2}
{3\sqrt{2}G_{F}\sin^2\theta M_{Z}^2} \; .
\end{equation}
where we find a relation between the $m_{\chi}$ and the $\xi$
angle for different values of $M_{Z'}$.

In Fig. 1, the region between the horizontal line and the
`parabolic'curves give the possible values for the $m_{\chi}$ and
the $\xi$ angle. For example, if $\xi = \pi/4$ then the $m_{\chi}=
1728$ GeV, 3458 GeV and, 6916 GeV for $M_{Z'} = 150$ GeV, 300 GeV
and 600 GeV respectively.

In conclusion, a relation between the masses and the mixing
angles of the gauge fields and the would-be
Goldstone bosons, for a theory with one additional neutral current,
is found, using the $R_{\xi}$ gauge fixing, and we are able to
recover the MSM constraints on the Higgs mass in the $\psi\approx
0$ case. The latter relation is valid for any extra neutral
current
and the differences depend on the coupling constants $g1$, $g2$
and the coefficients $a$, $b$ which define the electromagnetic
charge. This coefficients depend on the fermionic content of
the model and the cancellation of the anomalies.

\subsection*{Acknowledgements}
R.M. wishes to thank Professor A. Salam for the opportunity of
visiting the International Center for
Theoretical Physics and also COLCIENCIAS and The Third World
Academy of Sciencies for their partial financial support.\\
N.V. would like to thank the Universidad de Antioquia for their
support and the Fundaci\'on Mazda for their generous scholarship.

\newpage

\noindent FIG.\ 1 The region between the horinzontal line and the
`parabolic' curves gives the allowed values for the $m_{\chi}$
and the mixing angle $\xi$, for different values of $M_{Z'}$. The
dashed, the doted and, the dot - dashed curves correspond to the
$M_{Z'} = 150$ GeV, 300 GeV and 600 GeV respectively.

\end{document}